\documentclass[conference]{IEEEtranSPMB}

\IEEEoverridecommandlockouts
\ifCLASSINFOpdf
\else
\fi
\usepackage{amsmath,graphicx,soul,subcaption,balance}
\usepackage{balance}
\usepackage[numbers,sort&compress]{natbib}

\usepackage{parskip}
\usepackage{float} 

\usepackage{mathptmx}

\usepackage{multirow} 

\usepackage{color}
\usepackage{soul,mathrsfs}
\usepackage[utf8]{inputenc}
\usepackage[T1]{fontenc}
\usepackage{array}
\usepackage[hyphens, spaces, obeyspaces]{url}

\usepackage[papersize={8.5in,11in}, left=1in, right=1in, top=1in, bottom=1in]{geometry}
\newcolumntype{L}[1]{>{\raggedright\let\newline\\\arraybackslash\hspace{0pt}}m{#1}}
\newcolumntype{C}[1]{>{\centering\let\newline\\\arraybackslash\hspace{0pt}}m{#1}}
\newcolumntype{R}[1]{>{\raggedleft\let\newline\\\arraybackslash\hspace{0pt}}m{#1}}

\usepackage[singlelinecheck=false,justification=centering]{caption}
\DeclareCaptionLabelFormat{mylabel}{#1 #2.\hspace{0.7ex}}
\renewcommand{\thetable}{\arabic{table}}

\usepackage{xcolor,colortbl}
\definecolor{light-gray}{gray}{0.83}

\newcommand{\spmbtitlefont}{\fontsize{11.0pt}{11.00pt}\selectfont\bf\vspace{0.7em}}
\newcommand{\spmbauthorfont}{\fontsize{11.0pt}{11.0pt}\selectfont\vspace{0em}}
\newcommand{\subparagraph}{}
\usepackage[compact]{titlesec}
\titleformat{\section}
   {\center\normalfont\sc}{\thesection.}{0.7ex}{}
\titlespacing{\section}{0pt}{2ex}{1.5ex}
\titlespacing{\subsection}{0pt}{1.5ex}{1.2ex}
\titlespacing{\subsubsection}{0pt}{1ex}{0.9ex}
\makeatletter
\renewcommand*{\@seccntformat}[1]{\csname the#1\endcsname .\hspace{0.7em}}
\makeatother

\usepackage{fancyhdr,lastpage}
\fancypagestyle{firststyle}
{
}
\title{\spmbtitlefont Racial Disparities in Pulse Oximetry Cannot Be Fixed With Race-Based Correction
{\vspace{-2.4\baselineskip}
}
}
    \author{\spmbauthorfont\IEEEauthorblockN{
    Neal Patwari, Di Huang, and Kiki Bonetta-Misteli
    }
    \vspace{0.9em}
    \IEEEauthorblockA{\spmbauthorfont 
        McKelvey School of Engineering, Washington University in St.\ Louis, Missouri, USA \\
        \{npatwari, di.huang,  fbonetta-misteli\}@wustl.edu
    }
}

\newcommand{\PaperTitleSummary}{N.\ Patwari, et al.: Racial Disparities in Pulse Oximetry Cannot Be Fixed ...}

\begin{document}

\IEEEaftertitletext{}
\maketitle

\begin{abstract}
Studies have shown pulse oximeter measurements of blood oxygenation have statistical bias that is a function of race, which results in higher rates of occult hypoxemia, i.e., missed detection of dangerously low oxygenation, in patients of color.  This paper further characterizes the statistical distribution of pulse ox measurements, showing they also have a higher variance for patients racialized as Black, compared to those racialized as white.  We show that no single race-based correction factor will provide equal performance in the detection of hypoxemia.  The results have implications for racially equitable pulse oximetry.
\end{abstract}

\begin{keywords}
racism, health care, inequity, occult hypoxemia
\end{keywords}

\IEEEpeerreviewmaketitle    
\thispagestyle{firststyle}  
\section{Introduction} \label{sec:introduction}

Multiple studies of pulse oximetry have reported on the racial bias of pulse oximeter measurements over the past 30 years \cite{jubran1990reliability,feiner2007dark,sjoding2020racial,andrist2022association}.  These reports show average pulse oximeter (SpO$_2$) measurements are biased higher for patients racialized as Black, particularly at low blood oxygenation levels.  Since pulse oximeters are used to detect hypoxemia, i.e., critically low blood oxygenation, the higher positive bias in SpO$_2$ for Black patients results in more frequent occult (missed detection of) hypoxemia, as compared to white patients.  The reported odds ratio for occult hypoxemia for Black, compared to white, patients is 2.2 \cite{andrist2022association}, 1.4 \cite{burnett2022selfreported}, and ``nearly three'' \cite{sjoding2020racial}.

This discrepancy influences life-saving care. Those infected with COVID-19 are advised to go to an emergency department only if their pulse ox readings are below 90\% \cite{moran2020popular}, and Medicare reimbursement rules are based on oxygenation percentage.  As a result, white patients are more able to obtain care than Black patients when faced with equally dangerous low blood oxygenation.  For those hospitalized with COVID-19, ``it is possible that unreliable measurements of the oxygen saturations have contributed to increased mortality reported in Black patients'' \cite{tobin2022pulse}. By hiding a patient's hypoxemia behind acceptable oxygenation readings, the pulse oximeter contributes to the racialized and gendered phenomenon that Sasha Ottey calls ``health care gaslighting'', in which a patient in a medical crisis is told that they are not in danger.

The word ``bias'' has two meanings: \textit{statistical} bias, and \textit{differential} assessment by group.  In the case of pulse oximetry, we observe both.  Arterial oxygen saturation estimates from pulse oximeters themselves have a statistical bias, that is, the expected value of the pulse oximeter measurement is different from the arterial oxygenation saturation.  This difference is called the \textit{statistical bias}.  The racial discriminatory impact to patients stems, in part, from the racial differences in the statistical bias.  That is, the bias is more severe in patients racialized as Black vs.\ white, which then results in disparate rates of care for hypoxemia.  This performance difference in pulse oximetry has been observed since at least 1990 \cite{jubran1990reliability}, but has not been fixed.   

If the racial difference in pulse oximetry was only in the mean of the measurements, this could lead practitioners to use a race-based correction factor.  In this case, the statistical bias as a function of race would be subtracted out, as a function of the patient's racial group, to produce bias-corrected measurements.  Indeed, this is suggested for patient care \cite{philip2021racial}.  There are many problems, historically, with race-based correction factors, which have been used in the US to justify treating Black patients as less than human and as less in need of medical care \cite{vyas2020hidden}. Race-based correction is also problematic because people may be in more than one racial group; and due to the social construction of race, skin color is not the same as race.  Beyond these very practical problems, in this paper, we also quantify the argument that \textit{no race-based correction factor will lead to equal performance in hypoxemia detection between patients racialized as Black and white}.

Using a large patient data set \cite{pollard2018eicu}, this paper goes beyond the Black/white racial differences in statistical bias in two ways:
\begin{enumerate}
    \item We address the racial differences in the shape of the distribution, in particular the variance, of pulse oximeter measurements across five racial groups. Clinicians have asked for this additional information \cite{philip2021racial}.
    \item We show how the racial differences in statistical distribution of pulse oximeter values leads to hypoxemia detection differences that cannot be corrected simply by removing the statistical bias as a function of race.
\end{enumerate}
Together, these two contributions help to motivate and inform the effort to design racially equitable pulse oximeters.  We present in Section \ref{sec:discussion} a discussion of the impact of the analysis on future research.

\section{Data} \label{sec:data_source}

We use the eICU Collaborative Research Database (eICU-CRD) in our retrospective analysis \cite{pollard2018eicu}.  It is the larger of the two data sets used in the influential Sjoding et al.\ study \cite{sjoding2020racial} and is publicly available.  In this section, we describe the database and how we extract pulse oximeter  and arterial blood gas measurements.

The eICU-CRD has anonymized records from over 139,000 unique patients during their critical care in an ICU between 2014 and 2015.  The eICU-CRD is unique because of its size --- 208 hospitals contributed, and it contains a huge stream of data from each patient associated with their care and physiological measurements recorded during their stay, including the pulse oximeter and blood gas measurements we use in this paper. 
In summary, we extract from the eICU-CRD each arterial blood gas measurement (SaO$_2$), and then find the pulse oximeter value (SpO$_2$) for the same patient that is measured closest in time.  If an SpO$_2$ was measured within $T=10$ minutes of their SaO$_2$ measurement, we store the (SaO$_2$, SpO$_2$) pair for use in our study, along with the patient race/ethnicity.  While blood oxygenation is not constant in clinical settings within  $10$ minutes \cite{okunlola2022pulse}, our duration matches that of Sjoding et al.\ \cite{sjoding2020racial}.  Further, although not reported here, we do not see significant differences in our results when we apply $T=5$ or $T=20$ minutes.

For more detail on this process, we note the eICU-CRD contains multiple tables, linked to one individual by a unique and random identifier, \verb|patientUnitStayID|, ``which uniquely identifies a single ICU stay of a patient'' \cite{pollard2018eicu}. Measurements are timestamped, in minutes after the time of the patient's ICU admission. We extract 335k measurements of SaO$_2$ from 73k unique patients from the \verb|lab| table, measured via arterial blood gas (ABG) test. We extract 140M measurements of SpO$_2$ from 196k unique patients from the following tables: \verb|vitalPeriodic| (5-min median from the bedside pulse oximeter), \verb|physicalExam| (only the `current value' record), and \verb|nurseCharting| (entered by a nurse from a bedside reading).

Each patient  has exactly one race or ethnicity entry in the `patient' table, with options are limited to: white, African American or Black, Hispanic, Asian, Native American, or ``other/unknown'', presumably pulled from the patient's medical record \cite{pollard2018eicu}. Race and ethnicity itself does not determine a person's skin pigmentation, as race is a social construct. We discuss the impact of these factors in Section \ref{sec:discussion}.

\begin{figure}[htbp]
         \centering
         \includegraphics[width=0.95\columnwidth]{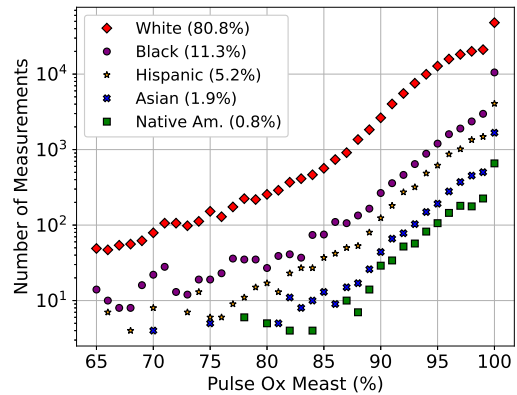}
         \caption{Number of measurements by race and SpO$_2$ value (and total \% of data by race).}
         \label{fig:numMeasts}
\end{figure}

The result of the data extraction is that we have 218,000 pairs of (SaO$_2$, SpO$_2$) that are measured within 10 minutes of each other.  Pairs from white patients are 80.8\% of the dataset, which is skewed more white than the US population as a whole. Further, the number of data pairs in the dataset reduces as the SpO$_2$ value decreases. Figure \ref{fig:numMeasts} shows the number of data pairs by race and SpO$_2$ value.  At lower SpO$_2$ values, and for less well-represented racial categories, we have less data with which to characterize pulse ox performance.  For example, for SpO$_2$ values $<$ 87\%, we do not have more than 10 data pairs from Native American patients, and data for Asian patients is often fewer than 10 pairs.  Thus, in this paper, when displaying results for all five available race/ethnicity categories, we limit the presentation to SpO$_2$$\ge$ 87\%.

\section{Methods and Results} \label{sec:results}

\subsection{Racial Difference in Statistical Bias} 

From the (SpO$_2$, SaO$_2$) pairs, we first validate prior work that identified a higher statistical bias in SpO$_2$ values for Black patients as compared to white patients.  We use ``SpO$_2$ error'' to refer to SpO$_2$ $-$ SaO$_2$.  The statistical bias is the expected value of the SpO$_2$ error.  We use the average error over all data pairs and display this average vs.\ racial category in Table \ref{t:race_all_bias}.  To provide a robust estimate less influenced by very low and high errors, we also show the median difference.  The statistical bias is consistently positive for all groups (1.71\%), which indicates pulse oximeters are reporting values higher on average than the corresponding SaO$_2$.

In our data, the bias is statistically significantly higher for Black patients (2.60\%) and Asian patients (2.47\%) vs.\ white patients (1.58\%).  We use a two-sided Welch's t-test to compare the data from each group to the white group; this test of difference in mean does not assume that the variances are identical across groups.  The test gives $p$-values less than $10^{-13}$ for data from Black and Asian patients, but $p>0.05$ for data from patients racialized as Hispanic or Native American.

\begin{table}[tbh]
\centering
 \begin{tabular}{| l | c c | c |} 
 \hline
 & \multicolumn{2}{c|}{\textbf{Bias}} & \textbf{Probability of} \\
 \textbf{Race/Ethnicity} & \textbf{Median} & \textbf{Average} & \textbf{$|$SpO$_2$$-$SaO$_2$$| > 10$} \\ 
\hline
White & 1.00\% & 1.58\% & 0.0486 \\
Black & 1.70\% & 2.60\% *** & 0.0813 *** \\
Asian & 1.30\% & 2.47\% *** & 0.0537 \\
Hispanic & 1.20\% & 1.47\% & 0.0384\\
Native American & 1.40\% & 1.81\% & 0.0407\\
 \hline
All Data & 1.00\% & 1.71\% & 0.0518\\
\hline \end{tabular}
 \caption{Pulse Ox Bias: Median, Average, \& Probability of Large ($>10$) Error, vs.\ Assigned Patient Race/Ethnicity.  ***$p<0.001$, from Welch's t-test used to test difference of mean compared to white patients.}
\label{t:race_all_bias}
\end{table}

\begin{figure}
         \centering
         \includegraphics[width=0.95\columnwidth]{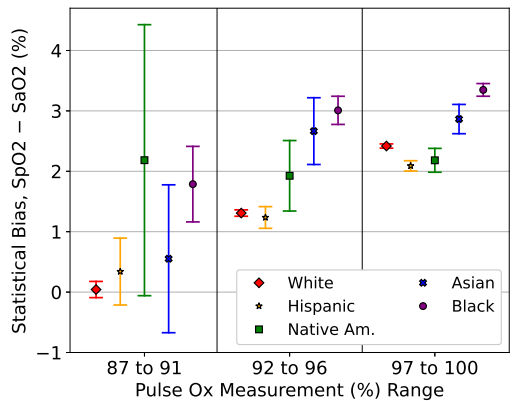}
         \caption{Statistical bias of SpO$_2$ values: 95\% confidence interval on mean of SpO$_2$ $-$ SaO$_2$, within three ranges of SpO$_2$ values, by race.}
         \label{fig:bias}
\end{figure}

Further, as observed across several papers in the literature \cite{feiner2007dark,andrist2022association}, the statistical bias is a function of blood oxygenation percentage.  How do the racial differences in average error change as a function of the measured SpO$_2$ value?  To show this, we divide the data into three ranges of SpO$_2$: Low [87 -- 91]; Medium [92 -- 96];  High [97 --100].  Each range is inclusive, and the lower limit of 87 is because of the low count of data points for Asian and Native American patients below 87, as displayed in Fig.\ \ref{fig:numMeasts}. The 95\% confidence interval for the statistical bias, i.e., the mean of SpO$_2$$-$SaO$_2$, is plotted in Fig.\ \ref{fig:bias} as a function of SpO$_2$ range and race.  In general, for all groups, the statistical bias decreases as the measured SpO$_2$ decreases. However, \textit{the racial disparity in the statistical bias increases as SpO$_2$ decreases}. At high SpO$_2$, the SpO$_2$ bias is 0.93 higher for Black patients than it is for white patients. In comparison, at low and medium SpO$_2$, the SpO$_2$ bias is about 1.70 higher for Black patients than it is for white patients.

Low numbers of data points for Asian and Native American patients results in wider confidence intervals at all SpO$_2$ ranges, particularly in the lowest SpO$_2$ range.  However,  SpO$_2$ bias for data from patients classified as Asian is generally higher than that from white patients and lower than that from Black patients. The statistical bias for patients racialized as Hispanic and Native American is sometimes higher than and sometimes lower than that of patients racialized as white.

\subsection{Racial Difference in Distribution Shape}

In this subsection, we investigate the shape of the statistical distribution of the SpO$_2$ error and see that there are differences in how wide they are for different racial groups.

\begin{figure*}[tbp]
     \centering
     \begin{subfigure}[b]{0.45\textwidth}
         \centering
         \includegraphics[width=\textwidth]{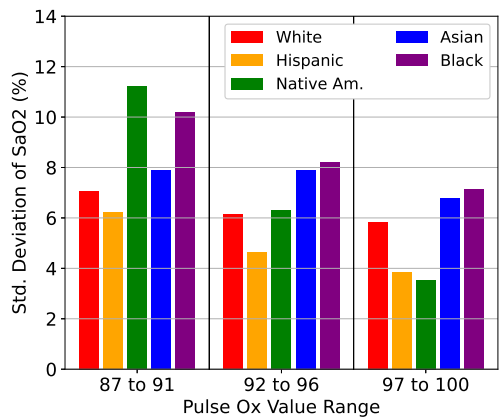}
         \caption{Estimated standard deviation of SaO$_2$ values  within three ranges of SpO$_2$ values, by race.}
         \label{fig:stdev}
     \end{subfigure}
     \hfill
     \begin{subfigure}[b]{0.45\textwidth}
         \centering
         \includegraphics[width=\textwidth]{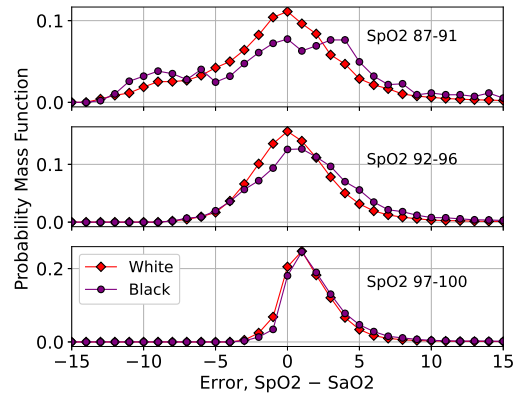}
         \caption{Probability mass function (pmf) of error, SpO$_2$ $-$ SaO$_2$, within three ranges of SpO$_2$ values, for data from patients classified as white and Black.}
         \label{fig:pmf}
     \end{subfigure}
        \caption{Distribution of SpO$_2$ $-$ SaO$_2$ for three ranges of SpO$_2$, by race.}
        \label{fig:four_graphs}
\end{figure*}

We plot in Figure \ref{fig:stdev} the standard deviation of the SpO$_2$ error, for three ranges of SpO$_2$ values and each race.  We observe that the standard deviation decreases as SpO$_2$ value increases.  As a function of race, the standard deviation of SpO$_2$ error is consistently higher for Asian and Black patients vs.\ white patients.  In fact, a one-sided F-test on the variance shows $p < 0.001$ for the variance of SpO$_2$ error from Asian patients or from Black patients, compared to white patients.  

For more information about the shape of these distributions, 
we plot the probability mass functions (pmfs) for each range of SpO$_2$ and for data from patients racialized as Black and as white, the two racial groups with sufficient data to plot pmfs.  With the exception of the high SpO$_2$ range, Figure \ref{fig:pmf} shows SpO$_2$ error for white patients is narrower than for Black patients.

In the highest SpO$_2$ group, the high statistical bias may artificially make the variance lower for all racial groups, since the maximum oxygenation is 100\%.  This can be seen in Figure \ref{fig:pmf} in the bottom subplot (corresponding to the high SpO$_2$ range) in which the negative tail of the pmf is significantly compressed compared to the heavy positive tail.  Since the SpO$_2$ error is defined as SpO$_2$$-$SaO$_2$, the SpO$_2$ is in the range 97-100, and the SaO$_2$ can be at most 100, there is much more opportunity for the error to be positive rather than negative.

In low and medium SpO$_2$ ranges, the pmf plots in Figure \ref{fig:pmf} show the significantly heavier tails of the SpO$_2$ error for patients racialized as Black vs.\ those racialized as white.  For Black patients, very large errors of 10 or more have relatively high probabilities, as compared to white patients.  We compute and show the probabilities of errors larger than 10, for each racial category, in the rightmost column of Table \ref{t:race_all_bias}.  Patients racialized as Black have a 67\% higher chance of a large SpO$_2$ error compared to patients racialized as white.  Using a one-sided test of equal proportions, via a normal approximation since the number of data points is large \cite{prins2012nist}, we see that this difference in large error probability is statistically significant for Black compared to white patients with $p< 0.001$.  The data from patients racialized as Asian has $p=0.068$. 

\subsection{Hypoxemia Detection Performance}

How do these distributions of SpO$_2$ impact detection of hypoxemia?  Hypoxemia is defined has having a arterial oxygenation saturation less than 88\% \cite{sjoding2020racial}. We use SaO$_2$ as a gold standard for arterial oxygenation saturation measurement, as before, in our analysis of the errors of SpO$_2$. Thus we label any data pair as true hypoxemia when it has an SaO$_2$ $< 88$\%.  We study a detector based solely on the single SpO$_2$ measurement taken within 10 minutes of the SaO$_2$ measurement, and our question is: what is the performance of a hypoxemia detector that uses one SpO$_2$ measurement as its input?  We consider a null hypothesis $H_0$: ``does not have hypoxemia'' and an alternate hypothesis $H_1$: ``has hypoxemia''; and a detector that uses a single threshold $\gamma$: it decides $H_1$ if the SpO$_2$ is less than $\gamma$, and decides $H_0$ if the SpO$_2$ is higher than $\gamma$. 

The detector threshold $\gamma$ becomes the parameter that can be tuned to trade off performance between the two types of detection errors:

\vspace{-0.1in}
\begin{enumerate}
    \item Type I error or ``false alarm': the patient does not have hypoxemia, but SpO$_2$$<\gamma$ and thus the detector decides $H_1$.
    \item Type II error or ``missed detection'': the patient \textit{has} hypoxemia, but SpO$_2$$>\gamma$ and  the detector decides $H_0$ --- also called ``occult hypoxemia''.
\end{enumerate}

\vspace{-0.1in}
Note that $\gamma$ does not need to be the same as the definition of hypoxemia (88\%).  

\begin{figure*}
         \centering
         (a) \includegraphics[width=0.9\columnwidth]{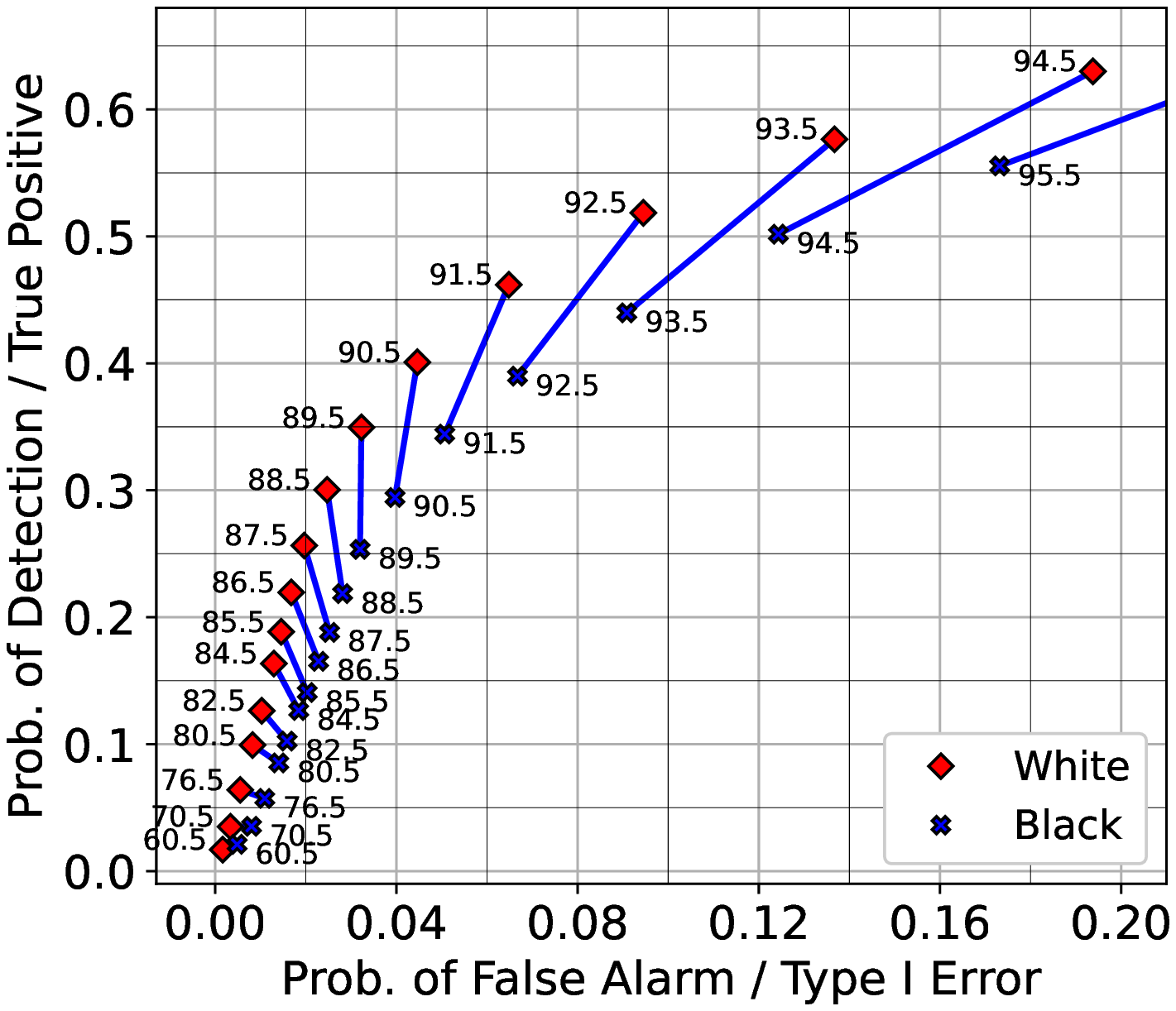} (b) \includegraphics[width=0.9\columnwidth]{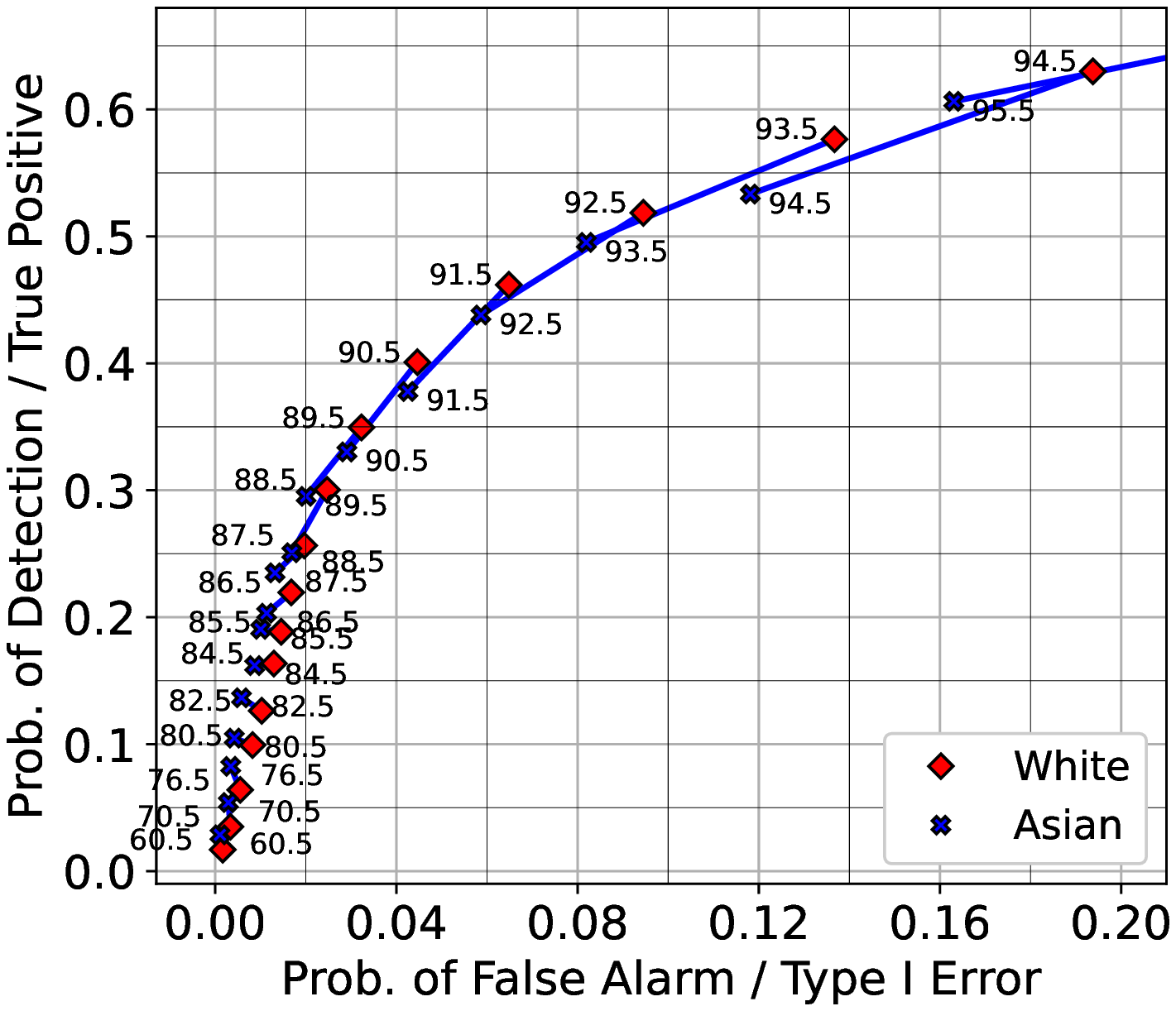}
         \caption{Probability of correct detection vs.\  probability of false alarm, for data from patients racialized as (a) Black vs.\ white, and (b) Asian vs.\ white, as a function of threshold $\gamma$. Ideal performance is at the top left corner.  Hypoxemia detection performance for same $\gamma$ is connected (\textcolor{blue}{\textbf{---}}); no single threshold results in equal performance.  No race-conscious thresholds can achieve equal  performance between Black and white patients, but they can achieve similar results between Asian and white patients.}
         \label{fig:roc}
\end{figure*}

Possible detector performance points are plotted for a wide range of thresholds $\gamma$, and for data from patients racialized as Black and white, in Fig.\ \ref{fig:roc}(a).  Each performance point is labelled with the $\gamma$ threshold value used to obtain it, and since SpO$_2$ values are all integers, we avoid confusion by setting $\gamma$ values to be halfway between integers, i.e., ending in ``.5''.  The performance gap between Black and white patients is shown by connecting the points with a blue line.  

Once can observe that no single threshold results in equal performance for Black and white patients. For example, $\gamma = 88.5$ results in a probability of detection of 30\% and probability of false alarm of 2.5\% for white patients; while the same probabilities are 22\% and 2.8\% for Black patients.   We emphasize the point that \textit{no single threshold can achieve the same performance point (probabilities of detection and false alarm) for both Black and white patients}.  

Further, \textit{no ``race-based'' correction will be able to equalize detection performance}.  Medical algorithms with different settings as a function of race are controversial and typically lead to better outcomes for white patients as compared to patients of color \cite{vyas2020hidden}. But even if we attempt to detect hypoxemia with different race-dependant thresholds with the goal of reducing disparities, we cannot achieve equal performance points.  For example, we might use $\gamma = 88.5$ (as described above) for patients racialized as white and $\gamma= 90.5$ for patients racialized as Black.  The $\gamma= 90.5$ for Black patients allows the probability of detection to be increased to 29.4\%, which would be nearly the same probability of detection as for white patients using $\gamma = 88.5$.  However, the probability of false alarm would still be disparate: 2.5\% and 4.0\% for patients racialized as white and Black, respectively.

As a comparison, we look at the same performance points for patients racialized as Asian in Figure \ref{fig:roc}(b).  Similar to the above case, no single threshold $\gamma$ produces identical detection performance for patients racialized as Asian vs.\ white.  However, the points are nearly on the same curve.  If we did consider a race-based correction factor, for example, subtracting 1.0 from the SpO$_2$ measurements for patients racialized as Asian, or equivalently increasing $\gamma$ by 1.0 for Asian patients, we could make a race-based correction factor for $\gamma$ for Asian patients that allows the detector to approximately match the hypoxemia detection performance for patients racialized as Asian and white.  In Figure \ref{fig:roc}(b), one can see that for $\gamma$ in the 87.5 to 92.5 range for white patients, and $\gamma$ in the 88.5 to 93.5 range for Asian patients, performance points are close and approximately on the same curve.  Further study is required for conclusive results due to having 20 times less data for Asian compared to white patients. Note that such a correction would not fix the discrimination against Black patients.

\section{Discussion and Future Research} \label{sec:discussion}

In Section \ref{sec:results}, we validate that there is a statistical bias in SpO$_2$ measurements that differs significantly by racial group.  If this were the only racial disparity in SpO$_2$ measurements, one could in theory have a race-based correction using different thresholds for each racial/ethnic category to achieve identical detection performance.  But this is not the only disparity between patients racialized as Black and white.  The significantly higher standard deviation and heavier tails of the SpO$_2$ distribution show that SpO$_2$ has larger errors for patients racialized as Black compared to white.  Essentially, SpO$_2$ for Black patients is \textit{both} less accurate and less precise than for white patients.  Correcting for a known statistical bias, as suggested in \cite{philip2021racial}, is insufficient to achieve equal performance for patients racialized as Black vs.\ white. Fixing the racial disparities in performance of pulse oximeters will require addressing both the mean and the variance of the SpO$_2$ errors. The literature about pulse oximeter disparities primarily addresses the mean, but not the variance. 

To discuss the difference in variance, consider the impact of melanin on pulse ox measurements.  In the visible light range, melanin is the main absorbent chromophore in human skin, while hemoglobin dominates absorbance in the blood.
The absorption of each is represented by their \textit{extinction spectra}, e.g., as in Figures 1.3 \& 1.4 of \cite{wang2012biomedical}. Typical pulse oximeters use two LEDs, one in the red visible light range (\textasciitilde650 nm) and one in the infrared range (\textasciitilde900 nm). The degree to which light is attenuated is proportional to the concentration of melanin in the skin at the recording site, and red light is particularly attenuated by the presence of melanin. 

One way in which the variance of SpO$_2$ is impacted is due to this signal attenuation. Because of the attenuation of light due to melanin, the signal to noise ratio (SNR) of the pulse oximeter light measurements is reduced, particularly at the red wavelength. As the blood oxygenation estimate is a function of light measurements at these two wavelengths, the lower SNR of the measurements will lead to higher variance estimates for those with more melanated skin.
    
Another way in which the variance of SpO$_2$ is lower among patients racialized as white may be due to the particular way in which race has been socially constructed in the US over its 400 year history. Race was constructed during slavery, in a way that ``allowed white men to profit from their sexual assaults on Black women'' \cite{roberts2021race} by enslaving people if they were descendant from a person and their enslaver. After the end of slavery, laws enforced white supremacy with the \textit{one-drop rule}, that any person with any degree of Black ancestry would be considered Black \cite{brook2019accident}. This construction also ensures that people with a wide variety of melanin levels are racialized as Black.  If the statistical bias in SpO$_2$ is a function of melanin level, then SpO$_2$ values for those racialized as Black will include a wide variety of SpO$_2$ biases; grouped together this will make in the SpO$_2$ distribution among Black patients wider, i.e., with a higher variance.

Since race is insufficient to determine skin color, and the eICU-CRD doesn't report it, we are unable to further explore the above mechanisms. It is essential that a future large pulse oximetry study quantify skin color in addition to recording race, as suggested in \cite{okunlola2022pulse}.  

Future studies must include larger numbers of patients of color to quantify the implications of skin color on pulse oximetry.  The high proportion of white patients in the eICU-CRD makes it difficult to analyze errors in Asian and Native American patients.  We note that pulse oximeters are currently approved by the US FDA even when their subject studies include very few people of color.  The FDA 510(k) process recommends, but does not require, that the testing subject population includes, ``at least 2 darkly pigmented subjects or 15\%'' of the total pool \cite{okunlola2022pulse}. If instead the FDA process had a clear requirement for an over-representation of people of color in the subject population, a pulse ox study could definitively surface bias, and possibly mechanisms to fix bias issues, prior to public use.

\section{Conclusion}

In pulse oximetry, it is painfully clear that, as Inioluwa Deborah Raji says, ``the data that we handle are human fates, not footnotes'' \cite{raji2020handle}.  If racial disparities had been addressed with new pulse oximeter designs starting when they were first published over three decades years ago \cite{tobin2022pulse}, we might not have had pulse oximetry contributing to the racial disparities in outcomes during the SARS-CoV-2 pandemic.  In this paper, we use the eICU-CRD  \cite{pollard2018eicu} data previously reported on by Sjoding et al.\ \cite{sjoding2020racial}.  We contribute to the discussion with an in-depth analysis of the statistical distributions of SpO$_2$ errors as a function of race.  We show that, in addition to a statistical bias that differs by race, that the SpO$_2$ error distribution is wider for patients racialized as Black vs.\ white. We further show that any hypoxemia detector, whether it includes race or not, cannot achieve identical hypoxemia detection performance between white and Black patients.  We argue that these results, in the context of the history of racialization in the US, provide specific guidance for the direction of future pulse oximetry research and testing.

\setlength{\bibsep}{2.5pt plus 0ex}
\footnotesize
\bibliographystyle{IEEEbibSPMB}
\bibliography{sample-base}

\end{document}